\documentclass[aip,jcp,preprint,showpacs,superscriptaddress,groupedaddress]{revtex4-1}

\usepackage{amsmath}
\usepackage{amssymb}
\usepackage{braket}
\usepackage{bm}

\begin{document}

\title{Pertubative corrections for Hartree-Fock-like Algebraic Bethe Ansatz analogue.}

\author{Jean-David Moisset}
\author{Laurie Carrier}
\author{Paul A. Johnson}
 \email{paul.johnson@chm.ulaval.ca}
 
 \affiliation{D\'{e}partement de chimie, Universit\'{e} Laval, Qu\'{e}bec, Qu\'{e}bec, G1V 0A6, Canada}

\date{\today}

\begin{abstract}
Rayleigh-Schr\"{o}dinger perturbation theory corrections are developed for an algebraic Bethe ansatz of individual electrons. Numerical results are ambiguous and would need either an orbital optimization or a configuration interaction singles solution to be satisfactory. Two different expressions are computed to be certain that results are correct.
\end{abstract}

\maketitle

\section{Introduction}
Strongly-correlated systems present a challenge in quantum chemistry. The dominant behaviour is not independent electrons and thus methods built in that framework do not perform well. Powerful methods exist but are generally either too difficult or too expensive, though substantial improvement is being made.\cite{chan:2011,booth:2013,holmes:2017,li:2018} A promising alternative is to consider wavefunctions built from weakly-correlated pairs of electrons, or geminals. This idea is not new,\cite{hurley:1953,coleman:1965,silver:1969,silver:1970,coleman:1997} though a strong renewed interest\cite{surjan:1999,kobayashi:2010,surjan:2012,neuscamman:2012,peter:2013,johnson:2013,stein:2014,boguslawski:2014a,boguslawski:2014b,boguslawski:2014c,tecmer:2014,henderson:2014a,henderson:2014b,shepherd:2014,bulik:2015,pastorczak:2015,henderson:2019,khamoshi:2019,henderson:2020,dutta:2020,harsha:2020,khamoshi:2020,khamoshi:2021,dutta:2021,marie:2021} has shown that these types of wavefunctions correctly describe many bond-breaking processes. In particular, the antisymmetric product of 1-reference orbital geminals (AP1roG), or equivalently pair coupled-cluster doubles (pCCD), scales like $\mathcal{O}(N^4)$ and dissociates Hydrogen chains almost perfectly. Their principal drawback is that they must be solved projectively rather than variationally to be feasible. 

We are building approaches for strongly-correlated electrons from exactly solvable models. In particular we are using the algebraic Bethe ansatz (ABA)\cite{bethe,faddeev:1981} solution to the reduced Bardeen-Cooper-Schrieffer Hamiltonian,\cite{bardeen:1957a,bardeen:1957b} which we refer to as Richardson-Gaudin (RG)\cite{richardson:1963,richardson:1964,richardson:1965,gaudin:1976,dukelsky:2004} states, as a mean-field geminal wavefunction.\cite{johnson:2020,fecteau:2020,fecteau:2021,johnson:2021} The ABA is an approach capable of solving a large class of models both in quantum mechanics,\cite{QISM_book} and in 2-dimensional classical statistical mechanics.\cite{baxter_sm} We are studying RG states as the general mean-field for pairs of electrons, the so-called antisymmetric product of interacting geminals (APIG), is intractable to compute whereas RG states have polynomial cost and may be improved upon systematically. In a previous contribution\cite{carrier:2020} we studied an ABA for individual electrons to gauge how well the general mean-field, in that case Hartree-Fock (HF), was reproduced. We demonstrated numerically that there is essentially no approximation at the mean-field level.

The purpose of this contribution is to develop perturbative corrections for the ABA for electrons. The final equations look quite similar to the M\o{}ller-Plesset (MP) corrections to Hartree-Fock, though the singles give a non-zero contribution.\cite{MEST} Finally, the numerical results are ambiguous so we will not further explore the ABA for individual electrons. To be certain, we derived two different expressions that gave consistent numerical results. More optimization parameters would be required, either in the form of an orbital optimization or a configuration interaction (CI) singles solution.  

In section \ref{sec:aba} we briefly review the ABA for individual electrons in a general spin-orbital basis before moving to an unrestricted basis and calculating reduced density matrix (RDM) and transition density matrix (TDM) elements. Section \ref{sec:AO_RSPT} presents second-order Rayleigh-Schr\"{o}dinger perturbation theory (RSPT) corrections computed in the original ``primitive'' basis, while section \ref{sec:MO_RSPT} presents RSPT expressions in the basis of ABA objects (the analogue of molecular orbitals). 

\section{ABA for individual electrons} \label{sec:aba}
The ABA provides the eigenvectors of the Hamiltonian
\begin{align} \label{eq:ABA_ham}
\hat{H}_{ABA} &= \sum_i \varepsilon_i a^{\dagger}_i a_i + g \sum_{ij} a^{\dagger}_i a_j
\end{align}
which describes an aufbau filling of the lowest single particle states $\{\varepsilon\}$ along with an isotropic scattering $g$ of electrons between each site. Obviously, this Hamiltonian may be solved exactly by diagonalization. The point here is to try to understand the ABA in a simple case so that we can use that information for RG, where the analogue of HF, APIG, is intractable. 

We employ spin-orbitals that have the usual structure,
\begin{align}
[a^{\dagger}_i, a_j]_+ = \delta_{ij}
\end{align}
for which $i$ and $j$ are a complete set of spin and orbital indices. The ABA is built from a set of objects in terms of a complex number $u$:
\begin{align}
a^{\dagger}(u) &= \sum_i \frac{a^{\dagger}_i}{u - \varepsilon_i} \\
a(u) &= \sum_i \frac{a_i}{u-\varepsilon_i}.
\end{align}
These objects have the structure
\begin{align}
[a^{\dagger}(u), a(v)]_+ = \frac{A(u) - A(v)}{u - v}
\end{align}
where the scalar function $A(u)$ is
\begin{align}
A(u) = \frac{1}{g} - \sum_i \frac{1}{u - \varepsilon_i}.
\end{align}
In the limit that $u$ and $v$ are the same, this is well defined, with the result
\begin{align}
[a^{\dagger}(u), a(v)]_+ = \frac{\partial A(u)}{\partial u}.
\end{align}

\subsection{ABA solution}
The point of the ABA is that eigenvectors of \eqref{eq:ABA_ham} are products of $a^{\dagger}(u)$ whose arguments satisfy a set of equations. Usually these equations are coupled, but for this particular case they are not. We denote the ABA states based on their \emph{rapidities} $\{u\}$ as
\begin{align}
\ket{ \{u\} } = a^{\dagger}(u_1) a^{\dagger}(u_2) \dots a^{\dagger}(u_{2M}) \ket{\theta}
\end{align}
where the vacuum state $\ket{\theta}$ is understood such that $a_j \ket{\theta} =a(u) \ket{\theta} = 0 $. We emphasize that there are $2M$ electrons in $2N$ spin-orbitals. The state $\{u\}$ is a Slater determinant of $2M$ electrons in the basis of $a^{\dagger}(u)$. To establish the action of \eqref{eq:ABA_ham} on this state, the strategy is to move $\hat{H}_{ABA}$ to the right until it destroys the vacuum. First, it is easy to confirm that
\begin{align}
[ \hat{H}_{ABA} , a^{\dagger}(u)] = u a^{\dagger}(u) - g A(u) \sum_i a^{\dagger}_i 
\end{align}
and after a little work
\begin{align} \label{eq:on-shell}
\hat{H}_{ABA} \ket{ \{u\}} = \sum_{a\in occ} u_a \ket{\{u\}}  - g \sum_{a \in occ} A(u_a) \sum_i a^{\dagger}_i \ket{ \{u\}_a}.
\end{align}
Thus we have an eigenvector provided that for each $a$
\begin{align} \label{eq:so-BAE}
A(u_a) = \frac{1}{g} - \sum_i \frac{1}{u_a - \varepsilon_i} = 0.
\end{align}
These are called the Bethe ansatz equations (BAE). Usually they are coupled, and must be solved for each ABA state. In this particular case, the BAE are decoupled, and hence one solution suffices \emph{for all} the ABA eigenvectors. The BAE \eqref{eq:so-BAE} have $2N$ solutions, but each eigenvector only contains $2M$. To be in line with HF, we will refer to the rapidities that are present in the ground state as \emph{occupied} and those that do not as \emph{virtual}. The summations over $a$ in equation \eqref{eq:on-shell} are of course over only the occupied rapidities. In this particular case, the BAE may be solved by diagonalizing the matrix
\begin{align}
\begin{pmatrix}
\varepsilon_1 + g & g & \dots & g \\
g & \varepsilon_2 + g & \dots & g \\
\vdots & \vdots & \ddots & \vdots \\
g & g & \dots & \varepsilon_{2N} + g
\end{pmatrix}
\rightarrow
\begin{pmatrix}
u_1 & 0 &  \dots & 0 \\
0 & u_2 & \dots & 0 \\
\vdots & \vdots & \ddots & \vdots \\
0 & 0 & \dots & u_{2N}
\end{pmatrix}
\end{align}
which is substantially more stable numerically.

\subsection{Scalar Products}
Scalar products are computed by the form factor approach.\cite{zhou:2002,faribault:2008,faribault:2010,GB:2011} Namely to evaluate $\braket{ \{v\}|a^{\dagger}_i a_j | \{u\}}$ move $a_j$ to the right until it destroys the vacuum. This yields a sum over terms that are evaluated as specific limits of a general scalar product formula. For arbitrary $\{u\}$ and $\{v\}$ we can evaluate the scalar product easily as the determinant of a $2M\times 2M$ matrix
\begin{align}
\braket{\{v\} | \{u\}} = \det_{ab\in occ} \left( \frac{A(u_a) - A(v_b)}{u_a-v_b} \right).
\end{align}
Henceforth, we take one of the sets, say $\{v\}$ to be on-shell, i.e. $A(v_b) = 0$ for each $b$ giving

\begin{align}
\label{eq:slavnov}
\braket{\{v\} | \{u\}} = \prod_{c\in occ} A(v_c) \det_{ab\in occ} \left(\frac{1}{u_a-v_b} \right).
\end{align}
In principal this expression could be simplified as the determinant of a Cauchy matrix has a simple closed form expression, but this is not productive as we will see briefly. All the required scalar products are specific limits of \eqref{eq:slavnov}. In particular, the norm is the limit when $\{u\}\rightarrow \{v\}$:
\begin{align}
\braket{ \{v\} | \{v\}} = \prod_{c\in occ} \frac{\partial A(v_c)}{\partial v_c}
\end{align}
since the only surviving terms come from indeterminate forms on the diagonal of the determinant.

To evaluate the 1-electron reduced density matrix (1-RDM) we will need the structure constants
\begin{align}
[a_j, a^{\dagger}(u)]_+ &= \frac{1}{u-\varepsilon_j} \\
[a^{\dagger}_i , a^{\dagger}(u)]_+ &= 0
\end{align}
to move $a_j$ to the right. The result is
\begin{align}
\braket{ \{v\} | a^{\dagger}_i a_j | \{u\}} &= \sum_{a\in occ} (-1)^{a-1} [a_j,a^{\dagger}(u_a)]_+ \braket{\{v\}|a^{\dagger}_i | \{u\}_a} \\
&= \sum_{a \in occ} \frac{\braket{\{v\} | \{u\}_{a\rightarrow i}} }{u_a - \varepsilon_i}
\end{align}
where $\{u\}_a$ denotes the set $\{u\}$ without $u_a$, and the state $\ket{\{u\}_{a\rightarrow i}}$ is $\ket{\{u\}}$ with $a^{\dagger}(u_a)$ replaced with $a^{\dagger}_i$ \emph{in the correct position} which removes the sign. Now, the original operators are the residues of the ABA operators 
\begin{align}
a^{\dagger}_i = \lim_{u \rightarrow \varepsilon_i} (u - \varepsilon_i ) a^{\dagger}(u)
\end{align}
which carries through to the scalar product \eqref{eq:slavnov}
\begin{align}
\braket{ \{v \} | \{u\}_{a\rightarrow \varepsilon_i}} &= \lim_{u_a \rightarrow i} (u_a - \varepsilon_i ) \braket{ \{v\} | \{u\}} \\
&= - \det G^a_i \prod_{c (\neq a) \in occ} A(u_c).
\end{align}
Here the matrix $G^a_i$ is the matrix from \eqref{eq:slavnov} with the $a$th column replaced:
\begin{align}
\begin{pmatrix}
\frac{1}{u_a - v_1} \\
\frac{1}{u_a - v_2} \\
\vdots \\
\frac{1}{u_a - v_M}
\end{pmatrix}
\rightarrow
\begin{pmatrix}
\frac{1}{\varepsilon_i - v_1} \\
\frac{1}{\varepsilon_i - v_2} \\
\vdots \\
\frac{1}{\varepsilon_i - v_M}
\end{pmatrix}.
\end{align}
Taking the limit $\{u\} \rightarrow \{v\}$ and normalizing we arrive at the 1-RDM elements
\begin{align}
\gamma_{ij} = \frac{\braket{ \{v\} | a^{\dagger}_i a_j | \{v\} }}{ \braket{\{v\} | \{v\} }}
= \sum_{a\in occ} \frac{1}{(v_a - \varepsilon_i)(v_a - \varepsilon_j)} \frac{1}{\frac{\partial A(v_a)}{\partial v_a}}.
\end{align}
The ABA operators represent electrons in orthogonal but not normal orbitals. The quantity
\begin{align}
\frac{\partial A(v_a)}{\partial v_a} = \sum_j \frac{1}{(v_a - \varepsilon_j)^2}
\end{align}
plays the role of the (diagonal) overlap matrix.

Similarly for two electron operators,
\begin{align}
\braket{ \{v\} | a^{\dagger}_i a^{\dagger}_j a_l a_k | \{u\}  } = \sum_{a < b \in occ} 
\begin{vmatrix}
\frac{1}{u_a - \varepsilon_k} & \frac{1}{u_a - \varepsilon_l} \\
\frac{1}{u_b - \varepsilon_k} & \frac{1}{u_b - \varepsilon_l}
\end{vmatrix}
\braket{ \{v\} | \{u\}_{ab\rightarrow ij} }.
\end{align}
and
\begin{align}
\braket{ \{v\} | \{u\}_{ab\rightarrow ij} } = \det G^{ab}_{ij} \prod_{c(\neq a,b) \in occ} A(u_c)
\end{align}
leads to the 2-electron reduced density matrix (2-RDM) elements
\begin{align}
\Gamma_{ijkl} &= \frac{\braket{ \{v\} | a^{\dagger}_i a^{\dagger}_j a_l a_k | \{v\} }}{ \braket{\{v\} | \{v\} }} \\
&= \sum_{a<b \in occ} 
\begin{vmatrix}
\frac{1}{v_a - \varepsilon_k} & \frac{1}{v_a - \varepsilon_l} \\
\frac{1}{v_b - \varepsilon_k} & \frac{1}{v_b - \varepsilon_l}
\end{vmatrix}
\begin{vmatrix}
\frac{1}{v_a - \varepsilon_i} & \frac{1}{v_a - \varepsilon_j} \\
\frac{1}{v_b - \varepsilon_i} & \frac{1}{v_b - \varepsilon_j}
\end{vmatrix}
\frac{1}{\frac{\partial A(v_a)}{\partial v_a} \frac{\partial A(v_b)}{\partial v_b}} \\
&= \gamma_{ik} \gamma_{jl} - \gamma_{il} \gamma_{jk}.
\end{align}
The 2-RDM factors into 1-RDM information as the wavefunction is factored into 1-electron wavefunctions. This extends to any order 
\begin{align}
\Gamma^{(N)}_{i_1 \dots i_N j_1 \dots j_N} &= \frac{\braket{ \{v\} | a^{\dagger}_{i_1} \dots a^{\dagger}_{i_N} a_{j_N} \dots a_{j_1} | \{v\} }}{\braket{ \{v\}|\{v\} }} \\
&= \sum_{a_1 < \dots a_N \in occ}
\begin{vmatrix}
\frac{1}{v_{a_1} - \varepsilon_{j_1}} & \dots & \frac{1}{v_{a_1} - \varepsilon_{j_N}} \\
\vdots & \ddots & \vdots \\
\frac{1}{v_{a_N} - \varepsilon_{j_1}} & \dots & \frac{1}{v_{a_N} - \varepsilon_{j_N}}
\end{vmatrix}
\begin{vmatrix}
\frac{1}{v_{a_1} - \varepsilon_{i_1}} & \dots & \frac{1}{v_{a_1} - \varepsilon_{i_N}} \\
\vdots & \ddots & \vdots \\
\frac{1}{v_{a_N} - \varepsilon_{i_1}} & \dots & \frac{1}{v_{a_N} - \varepsilon_{i_N}}
\end{vmatrix}
\frac{1}{\frac{\partial A(v_{a_1})}{\partial v_{a_1}} \dots {\frac{\partial A(v_{a_N})}{\partial v_{a_N}}}} \\
&= \begin{vmatrix}
\gamma_{i_1 j_1} & \dots & \gamma_{i_1 j_N} \\
\vdots & \ddots & \vdots \\
\gamma_{i_N j_1} & \dots & \gamma_{i_N j_N}
\end{vmatrix}.
\end{align}

Transition density matrix (TDM) elements are computed with the same approach and the results are even simpler. For single excitations, i.e. replacing one $v_a$ with $v_b$, denoted set-wise as $\{v\}^b_a$, there is only one non-zero contribution. All others are proportional to $A(v_b)$ without a vanishing denominator and hence are identically zero. The non-zero contributions from single excitations are
\begin{align}
\braket{ \{v\} | a^{\dagger}_i a_j | \{v\}^b_a } = \frac{1}{(v_b - \varepsilon_j)(v_a - \varepsilon_i)} \prod_{c (\neq a) \in occ} \frac{\partial A(v_c)}{\partial v_c}
\end{align}
and
\begin{align}
\braket{ \{v\} | a^{\dagger}_i a^{\dagger}_j a_l a_k | \{v\}^b_a } =  
\sum_{c (\neq a) \in occ}
\begin{vmatrix}
\frac{1}{v_b - \varepsilon_k} & \frac{1}{v_b - \varepsilon_l} \\
\frac{1}{v_c - \varepsilon_k} & \frac{1}{v_c - \varepsilon_l}
\end{vmatrix}
\begin{vmatrix}
\frac{1}{v_a - \varepsilon_k} & \frac{1}{v_a - \varepsilon_l} \\
\frac{1}{v_c - \varepsilon_k} & \frac{1}{v_c - \varepsilon_l}
\end{vmatrix}
\prod_{d (\neq a,c) \in occ} \frac{\partial A(v_d)}{\partial v_d} .
\end{align}

Double excitations have the only non-zero contributions 
\begin{align}
\braket{ \{v\} | a^{\dagger}_i a^{\dagger}_j a_l a_k | \{v\}^{cd}_{ab} } = 
\begin{vmatrix}
\frac{1}{v_c - \varepsilon_k} & \frac{1}{v_c - \varepsilon_l} \\
\frac{1}{v_d - \varepsilon_k} & \frac{1}{v_d - \varepsilon_l}
\end{vmatrix}
\begin{vmatrix}
\frac{1}{v_a - \varepsilon_i} & \frac{1}{v_a - \varepsilon_j} \\
\frac{1}{v_b - \varepsilon_i} & \frac{1}{v_b - \varepsilon_j}
\end{vmatrix}
\prod_{c (\neq a,b) \in occ} \frac{\partial A(v_c)}{\partial v_c}.
\end{align}

\subsection{Unrestricted Orbitals}\label{sec:U_ABA}
To arrive at restricted results, we work out the result in unrestricted orbitals before taking the limit where the $\alpha$ and $\beta$ elements are identical. The elements can be separated into two sets. It's equivalent to having two separate ABA set-ups:
\begin{align}
\hat{H}^{\alpha}_{ABA} &= \sum_i \varepsilon^{\alpha}_i a^{\dagger}_i a_i + g^{\alpha} \sum_{ij} a^{\dagger}_i a_j \\
\hat{H}^{\beta}_{ABA} &= \sum_i \varepsilon^{\beta}_i a^{\dagger}_i a_i + g^{\beta} \sum_{ij} a^{\dagger}_i a_j
\end{align}
with solutions the roots of the equations:
\begin{align}
A (v^{\alpha}) := \frac{1}{g^{\alpha}} - \sum_i \frac{1}{v^{\alpha} - \varepsilon^{\alpha}_i} &= 0 \\
A (v^{\beta}) := \frac{1}{g^{\beta}} - \sum_i \frac{1}{v^{\beta} - \varepsilon^{\beta}_i} &= 0.
\end{align}
 The scalar products simplify as the matrix entering in \eqref{eq:slavnov} becomes block diagonal. Specifically, it being understood that $\{v\} = \{v^{\alpha}\} \cup \{v^{\beta}\}$, we have 
\begin{align} \label{eq:UBAHF}
\left(  \hat{H}^{\alpha}_{ABA} + \hat{H}^{\beta}_{ABA} \right) \ket{\{v\}} = \left( \sum_{a \in occ} v^{\alpha}_a + v^{\beta}_a \right) \ket{\{v\}}
\end{align}
and the scalar product with an arbitrary ABA vector $\ket{\{u\}}$ is
\begin{align}
\braket{ \{v\} | \{u\} } \prod_{c\in occ} A(u^{\alpha}_c) A(u^{\beta}_c) \det_{ab \in occ} \left( \frac{1}{u^{\alpha}_a - v^{\alpha}_b} \right) 
\det_{a'b' \in occ} \left( \frac{1}{u^{\beta}_{a'} - v^{\beta}_{b'}} \right).
\end{align}
The norm of the on-shell state becomes
\begin{align}
\braket{ \{v\}|\{v\}} = \prod_{a\in occ} \frac{\partial A(v^{\alpha}_a)}{\partial v^{\alpha}_a} \frac{\partial A(v^{\beta}_a)}{\partial v^{\beta}_a}.
\end{align}
The 1-RDM elements are:
\begin{align}
\gamma^{\sigma}_{ij} = \frac{\braket{ \{v\} | a^{\dagger}_{i \sigma} a_{j \sigma} | \{v\}}}{\braket{ \{v\}|\{v\} }}= \sum_{a\in occ} \frac{1}{(v^{\sigma}_a - \varepsilon^{\sigma}_i)(v^{\sigma}_a - \varepsilon^{\sigma}_j)} 
\frac{1}{\frac{\partial A (v^{\sigma}_a)}{\partial v^{\sigma}_a}}
\end{align}
and the 2-RDM elements reduce to 1-RDM elements
\begin{align}
\Gamma^{\sigma \tau}_{ijkl} = \frac{\braket{ \{v\} | a^{\dagger}_{i \sigma} a^{\dagger}_{j \tau} a_{l \tau} a_{k \tau} | \{v\} }}{\braket{ \{v\} | \{v\} }} 
= \gamma^{\sigma}_{ik} \gamma^{\tau}_{jl} - \delta_{\sigma\tau} \gamma^{\sigma}_{il} \gamma^{\sigma}_{jk}.
\end{align}

As in the previous section, excited states are labelled by which occupieds are removed, and which virtuals are added, i.e. the state $\ket{\{v\}^{p\sigma}_{a\sigma} }$ is the state with $a\sigma$ occupied orbital replaced by the $p\sigma$ virtual orbital.

We will require transition density matrix elements, for which $\sigma$ denotes either spin, while $\bar{\sigma}$ denotes the opposite spin of $\sigma$. They are obtained in the same manner as for the spin-orbital basis:
\begin{align}
\braket{ \{v\} | a^{\dagger}_{i\sigma} a_{j\sigma} | \{v\}^{p\sigma}_{a\sigma} } = \frac{\braket{ \{v\}|\{v\} }}{(v^{\sigma}_p - \varepsilon^{\sigma}_j)(v^{\sigma}_a - \varepsilon^{\sigma}_i) \frac{\partial A(v^{\sigma}_a)}{\partial v^{\sigma}_a}} 
\end{align}

\begin{align}
\braket{ \{v\} | a^{\dagger}_{i\sigma} a^{\dagger}_{j\sigma} a_{l\sigma} a_{k\sigma} | \{v\}^{p\sigma}_{a\sigma} } =
\sum_{c (\neq a) \in occ}
\begin{vmatrix}
\frac{1}{v^{\sigma}_p - \varepsilon^{\sigma}_k} & \frac{1}{v^{\sigma}_p - \varepsilon^{\sigma}_l} \\
\frac{1}{v^{\sigma}_c - \varepsilon^{\sigma}_k} & \frac{1}{v^{\sigma}_c - \varepsilon^{\sigma}_l}
\end{vmatrix}
\begin{vmatrix}
\frac{1}{v^{\sigma}_a - \varepsilon^{\sigma}_i} & \frac{1}{v^{\sigma}_a - \varepsilon^{\sigma}_j} \\
\frac{1}{v^{\sigma}_c - \varepsilon^{\sigma}_i} & \frac{1}{v^{\sigma}_c - \varepsilon^{\sigma}_j}
\end{vmatrix}
\frac{\braket{\{v\} | \{v\} }}{\frac{\partial A(v^{\sigma}_a)}{\partial v^{\sigma}_a} \frac{\partial A(v^{\sigma}_c)}{\partial v^{\sigma}_c}}.
\end{align}
Strictly speaking, the summation excludes the term $c = a$, but this term would give zero contribution so will be included to simplify later summations. The last TDM element for single excitations is
\begin{align}
\braket{ \{v\} | a^{\dagger}_{i\sigma} a^{\dagger}_{j\bar{\sigma}} a_{l\bar{\sigma}} a_{k\sigma} | \{v\}^{p\sigma}_{a\sigma} } =
\frac{\braket{ \{v\} | \{v\} }}{(v^{\sigma}_p-\varepsilon^{\sigma}_k)(v^{\sigma}_a - \varepsilon^{\sigma}_i)\frac{\partial A( v^{\sigma}_a)}{\partial v^{\sigma}_a}} \gamma^{\bar{\sigma}}_{jl}.
\end{align}

Double excitations only couple with the ground state through two-electron operators:
\begin{align}
\braket{ \{v\} | a^{\dagger}_{i\sigma} a^{\dagger}_{j\sigma} a_{l\sigma} a_{k\sigma} | \{v\}^{p\sigma q\sigma}_{a\sigma b\sigma} } =
\begin{vmatrix}
\frac{1}{v^{\sigma}_p - \varepsilon^{\sigma}_k} & \frac{1}{v^{\sigma}_p - \varepsilon^{\sigma}_l} \\
\frac{1}{v^{\sigma}_q - \varepsilon^{\sigma}_k} & \frac{1}{v^{\sigma}_q - \varepsilon^{\sigma}_l}
\end{vmatrix}
\begin{vmatrix}
\frac{1}{v^{\sigma}_a - \varepsilon^{\sigma}_i} & \frac{1}{v^{\sigma}_a - \varepsilon^{\sigma}_j} \\
\frac{1}{v^{\sigma}_b - \varepsilon^{\sigma}_i} & \frac{1}{v^{\sigma}_b - \varepsilon^{\sigma}_j}
\end{vmatrix}
\frac{\braket{\{v\} | \{v\} }}{\frac{\partial A(v^{\sigma}_a)}{\partial v^{\sigma}_a} \frac{\partial A(v^{\sigma}_b)}{\partial v^{\sigma}_b}}
\end{align}
and
\begin{align}
\braket{ \{v\} | a^{\dagger}_{i\sigma} a^{\dagger}_{j\bar{\sigma}} a_{l\bar{\sigma}} a_{k\sigma} | \{v\}^{p\sigma q\bar{\sigma}}_{a\sigma b\bar{\sigma}} } =
\frac{\braket{ \{v\}| \{v\} }}{(v^{\sigma}_a - \varepsilon^{\sigma}_i)(v^{\bar{\sigma}}_b - \varepsilon^{\bar{\sigma}}_j)(v^{\sigma}_p - \varepsilon^{\sigma}_k) (v^{\bar{\sigma}}_q - \varepsilon^{\bar{\sigma}}_l) \frac{\partial A(v^{\sigma}_a)}{\partial v^{\sigma}_a} \frac{\partial A(v^{\bar{\sigma}}_b)}{\partial v^{\bar{\sigma}}_b}}.
\end{align}
All other TDM elements either vanish or will not connect through the Hamiltonian.

\section{PT expressions: Primitive basis} \label{sec:AO_RSPT}
We wish to solve the Coulomb Hamiltonian for molecules
\begin{align}\label{eq:Coulomb}
\hat{H}_C = \sum_{ij} h_{ij} \sum_{\sigma} a^{\dagger}_{i\sigma} a_{\sigma} + \frac{1}{2} \sum_{ijkl} V_{ijkl} \sum_{\sigma\tau} a^{\dagger}_{i\sigma}a^{\dagger}_{j\tau} a_{l\tau} a_{k\sigma}
\end{align}
where the 1- and 2-electron integrals are expressed in a basis $\{\phi\}$ 
\begin{align}
h_{ij} &= \int d\mathbf{r} \phi^*_i (\mathbf{r}) \left( - \frac{1}{2} \nabla^2 - \sum_I \frac{Z_I}{| \mathbf{r} - \mathbf{R}_I |} \right) \phi_j (\mathbf{r}) \\
V_{ijkl} &= \int d\mathbf{r}_1 d\mathbf{r}_2 \frac{\phi^*_i(\mathbf{r}_1)  \phi^*_j(\mathbf{r}_2)  \phi_k(\mathbf{r}_1)  \phi_l(\mathbf{r}_2)  }{| \mathbf{r}_1 - \mathbf{r}_2|}.
\end{align}
Using the ABA for electrons as a wavefunction ansatz, and expanding 
\begin{align}
\hat{H}_C \ket{\Psi_k} = E_k \ket{\Psi_k}
\end{align}
in Rayleigh-Schr\"odinger perturbation theory (RSPT), we obtain
\begin{align}
\hat{H}_C &= \hat{H}_0 + \lambda \hat{H}_1 \\
E_k &= E^{(0)}_k + \lambda E^{(1)}_k + \lambda^2 E^{(2)}_k + \dots \\
\ket{\Psi_k} &= \ket{\Psi^{(0)}_k} + \lambda \ket{\Psi^{(1)}_k} + \lambda^2 \ket{\Psi^{(2)}_k} + \dots.
\end{align}

Collecting powers of $\lambda$, we highlight the zeroth order problem
\begin{align}
\hat{H}_0 \ket{\Psi^{(0)}_k} = E^{(0)}_k \ket{\Psi^{(0)}_k}
\end{align}
which is identically the problem \eqref{eq:UBAHF}. The RSPT first-order correction to the wavefunctions are
\begin{align}
\ket{\Psi^{(1)}_k} = \sum_{l\neq k} \frac{\braket{ \Psi^{(0)}_l | \hat{H}_1 | \Psi^{(0)}_k }}{E^{(0)}_k - E^{(0)}_l} \ket{ \Psi^{(0)}_l}
\end{align}
and since our first order Hamiltonian is
\begin{align}
\hat{H}_1 = \hat{H}_C - \left( \hat{H}^{\alpha}_{ABA} + \hat{H}^{\beta}_{ABA} \right)
\end{align}
and for $k\neq l$, 
\begin{align}
\braket{\Psi^{(0)}_l |  \left( \hat{H}^{\alpha}_{ABA} + \hat{H}^{\beta}_{ABA} \right) | \Psi^{(0)}_k} = 0
\end{align}
we can instead write the first order correction as
\begin{align}
\ket{\Psi^{(1)}_k} = \sum_{l\neq k} \frac{\braket{ \Psi^{(0)}_l | \hat{H}_C | \Psi^{(0)}_k }}{E^{(0)}_k - E^{(0)}_l} \ket{ \Psi^{(0)}_l}.
\end{align}
The 2nd order energy is then
\begin{align} \label{eq:rspt2}
E^{(2)}_k = \sum_{l\neq k} \frac{\vert \braket{ \Psi^{(0)}_l | \hat{H}_C | \Psi^{(0)}_k } \vert^2}{E^{(0)}_k - E^{(0)}_l}.
\end{align}

The ABA excited states are simply ABA states with different ``occupied'' rapidities since the BAE \eqref{eq:so-BAE} do not couple rapidities. Like the case for Hartree-Fock, the only excited states that couple with the ground state through the Hamiltonian are single and double excitations. 

The parameters defining the ABA in an unrestricted basis are $\{\varepsilon^{\alpha}\},\{\varepsilon^{\beta}\},g^{\alpha},g^{\beta}$, and since the 2-RDM factors into the 1-RDM contributions $\gamma^{\alpha}_{ij},\gamma^{\beta}_{ij}$, as shown previously\cite{carrier:2020} the variational ground state energy expression is

\begin{align}
E^{ABA}_U &= E^{(0)}_0 + E^{(1)}_0 \\
&= \min_{\{\varepsilon^{\alpha}\},\{\varepsilon^{\beta}\},g^{\alpha},g^{\beta}} 
\sum_{ij} h_{ij} \left( \gamma^{\alpha}_{ij} + \gamma^{\beta}_{ij} \right) \nonumber \\
&+ \frac{1}{2} \sum_{ijkl} V_{ijkl} \left( \gamma^{\alpha}_{ik}\gamma^{\alpha}_{jl} - \gamma^{\alpha}_{il}\gamma^{\alpha}_{jk} + \gamma^{\alpha}_{ik}\gamma^{\beta}_{jl} + \gamma^{\beta}_{ij} \gamma^{\alpha}_{kl} + \gamma^{\beta}_{ik} \gamma^{\beta}_{jl} - \gamma^{\beta}_{il} \gamma^{\beta}_{jk} \right).
\end{align}
To arrive at a restricted expression, we take the $\alpha$ and $\beta$ parameters to be identical. This amounts to solving for one ABA Hamiltonian, and using states that use each rapidity twice (once for each spin projection). The resulting energy expression is
\begin{align} \label{eq:ABA_R}
E^{ABA}_R &= \min_{\{\varepsilon\},g} 2 \sum_{ij} h_{ij} \gamma_{ij} + \sum_{ijkl} V_{ijkl} (2\gamma_{ik}\gamma_{jl} - \gamma_{il}\gamma_{jk})
\end{align}
with the 1-RDM elements:
\begin{align}
\gamma_{ij} = \sum_{a\in occ} \frac{1}{(v_a - \varepsilon_i)(v_a-\varepsilon_j)} \frac{1}{\frac{\partial A(v_a)}{\partial v_a}}.
\end{align}

We will now work out the 2nd-order energy correction. Our strategy is to work out the unrestricted expression first, then take the limit that the two spin projections are identical to get the restricted version.

The ABA ground state couples with single and double excitations through the Hamiltonian \eqref{eq:Coulomb}, so \eqref{eq:rspt2} becomes
\begin{align}
E^{(2)}_{U,0} &= \sum_{\substack{a \in occ \\ p \in virt} } \sum_{\sigma} 
\frac{1}{v^{\sigma}_a - v^{\sigma}_p}
\frac{\vert \braket{\{v\} | \hat{H}_C | \{v\}^{p\sigma}_{a\sigma} } \vert^2}{\braket{\{v\}|\{v\}} \braket{ \{v\}^{p\sigma}_{a\sigma} | \{v\}^{p\sigma}_{a\sigma}} }   \nonumber \\
&+ \sum_{\substack{a<b \in occ \\ p<q \in virt} } \sum_{\sigma} 
\frac{1}{v^{\sigma}_a +v^{\sigma}_b - v^{\sigma}_p -v^{\sigma}_q}
\frac{\vert \braket{\{v\} | \hat{H}_C | \{v\}^{p\sigma q\sigma}_{a\sigma b\sigma} } \vert^2}{\braket{\{v\}|\{v\}} \braket{ \{v\}^{p\sigma q\sigma}_{a\sigma b\sigma} | \{v\}^{p\sigma q\sigma}_{a\sigma b\sigma}} }  \nonumber \\
&+  \sum_{\substack{ab \in occ \\ pq \in virt} } 
\frac{1}{v^{\alpha}_a +v^{\beta}_b - v^{\alpha}_p -v^{\beta}_q}
\frac{\vert \braket{\{v\} | \hat{H}_C | \{v\}^{p\alpha q\beta}_{a\alpha b\beta} } \vert^2}{\braket{\{v\}|\{v\}} \braket{ \{v\}^{p\alpha q\beta}_{a\alpha b\beta} | \{v\}^{p\alpha q\beta}_{a\alpha b\beta}} } .
\end{align}
With the TDM elements computed in section \eqref{sec:U_ABA}, this becomes
\begin{align}
E^{(2)}_{U,0} &= \sum_{\substack{a \in occ \\ p \in virt} } \sum_{\sigma}  \frac{1}{v^{\sigma}_a - v^{\sigma}_p} 
\frac{(D^{p\sigma}_{a\sigma})^2}{\frac{\partial A(v^{\sigma}_a)}{\partial v^{\sigma}_a} \frac{\partial A(v^{\sigma}_p)}{\partial v^{\sigma}_p}} \nonumber \\
&+ \frac{1}{2} \sum_{\substack{a<b \in occ \\ p<q \in virt} } \sum_{\sigma}
\frac{1}{v^{\sigma}_a +v^{\sigma}_b - v^{\sigma}_p -v^{\sigma}_q}
\frac{(W^{\sigma\sigma}_{abpq}-W^{\sigma\sigma}_{abqp})^2}{\frac{\partial A(v^{\sigma}_a)}{\partial v^{\sigma}_a} \frac{\partial A(v^{\sigma}_b)}{\partial v^{\sigma}_b} \frac{\partial A(v^{\sigma}_p)}{\partial v^{\sigma}_p} \frac{\partial A(v^{\sigma}_q)}{\partial v^{\sigma}_q}}
\nonumber \\
&+ \sum_{\substack{ab \in occ \\ pq \in virt} } 
 \frac{1}{v^{\alpha}_a +v^{\beta}_b - v^{\alpha}_p -v^{\beta}_q}
 \frac{(W^{\alpha\beta}_{abpq})^2}{\frac{\partial A(v^{\alpha}_a)}{\partial v^{\alpha}_a} \frac{\partial A(v^{\beta}_b)}{\partial v^{\beta}_b} \frac{\partial A(v^{\alpha}_p)}{\partial v^{\alpha}_p} \frac{\partial A(v^{\beta}_q)}{\partial v^{\beta}_q}}
\end{align}
where the coupling matrix elements for the singles are
\begin{align}
D^{p\sigma}_{a\sigma} = t^{\sigma}_{ap} + \sum_{c\in occ} \left(
\frac{W^{\sigma\sigma}_{acpc} - W^{\sigma\sigma}_{accp}}{\frac{\partial A(v^{\sigma}_c)}{\partial v^{\sigma}_c}}
+ \frac{W^{\sigma\bar{\sigma}}_{acpc}}{\frac{\partial A(v^{\bar{\sigma}}_c)}{v^{\bar{\sigma}}_c}}
\right)
\end{align}
and the integrals have been transformed directly:
\begin{align}
t^{\sigma}_{ap} &= \sum_{ij} \frac{h_{ij}}{(v^{\sigma}_a - \varepsilon^{\sigma}_i)(v^{\sigma}_p - \varepsilon^{\sigma}_j)} \\
W^{\sigma\tau}_{abpq} &= \sum_{ijkl} 
\frac{V_{ijkl}}{(v^{\sigma}_a - \varepsilon^{\sigma}_i)(v^{\tau}_v-\varepsilon^{\tau}_j)(v^{\sigma}_p-\varepsilon^{\sigma}_k)(v^{\tau}_q-\varepsilon^{\tau}_l)}.
\end{align}
The restricted expression is obtained in the limit that the two sets of parameters are identical. Specifically,
\begin{align}\label{eq:resPT1}
E^{(2)}_{R,0} &= 2 \sum_{a\in occ} \sum_{p \in virt} \frac{1}{v_a -v_p} \frac{(D^p_a)^2}{\frac{\partial A(v_a)}{\partial v_a}\frac{\partial A(v_p)}{\partial v_p}} \nonumber \\
&+ \sum_{ab\in occ} \sum_{pq \in virt} \frac{1}{v_a + v_b - v_p -v_q} \frac{W_{abpq}(2W_{abpq}-W_{abqp})}{\frac{\partial A(v_a)}{\partial v_a}\frac{\partial A(v_b)}{\partial v_b}\frac{\partial A(v_p)}{\partial v_p}\frac{\partial A(v_q)}{\partial v_q}}
\end{align}
where the matrix element for the singles contribution is
\begin{align}
D^p_a = t_{ap} + \sum_{b\in occ} \frac{2W_{abpb}-W_{abbp}}{\frac{\partial A(v_b)}{\partial v_b}}
\end{align}
and the integrals are
\begin{align}
t_{ap} &= \sum_{ij} \frac{h_{ij}}{(v_a-\varepsilon_i)(v_p - \varepsilon_j)}  \\
W_{abpq} &= \sum_{ijkl} \frac{V_{ijkl}}{(v_a - \varepsilon_i)(v_b - \varepsilon_j)(v_p - \varepsilon_k)(v_q - \varepsilon_l)}.
\end{align}
The expression \eqref{eq:resPT1} is identical to the restricted MP2 expression \cite{MEST} except that the orbitals are not normalized and there are non-zero contributions from single excitations.

\section{PT expressions: ABA basis} \label{sec:MO_RSPT}
Expression \eqref{eq:resPT1} was computed using the Hamiltonian \eqref{eq:Coulomb} in the primitive basis, using transition density matrix elements computed with the ABA. The integrals $h_{ij}$ and $V_{ijkl}$ end up being transformed directly. To be certain that \eqref{eq:resPT1} is correct, we can compute the 2nd order perturbation in another manner, by transforming \eqref{eq:Coulomb} to the basis of ABA quasiparticles (the equivalent of molecular orbitals) and using particle-hole excitations to generate the relevant excited states. Notice that with the \emph{complete} set of solutions to the BAE we get a linear transformation of the creation operators:
\begin{align}
\begin{pmatrix}
a^{\dagger} (v^{\sigma}_1) \\
\vdots \\
a^{\dagger} (v^{\sigma}_N)
\end{pmatrix} =
\begin{pmatrix}
\frac{1}{v^{\sigma}_1 - \varepsilon^{\sigma}_1} & \dots & \frac{1}{v^{\sigma}_1 - \varepsilon^{\sigma}_N} \\
\vdots & \ddots & \vdots  \\
\frac{1}{v^{\sigma}_N - \varepsilon^{\sigma}_1} & \dots & \frac{1}{v^{\sigma}_N - \varepsilon^{\sigma}_N}
\end{pmatrix}
\begin{pmatrix}
a^{\dagger}_{1\sigma} \\
\vdots \\
a^{\dagger}_{N \sigma}
\end{pmatrix}
\end{align}
or 
\begin{align} \label{eq:trans}
a^{\dagger} (\textbf{v}^{\sigma}) = C^{\sigma} \textbf{a}^{\dagger}_{\sigma}.
\end{align}
The matrix $C$ has a known explicit inverse,
\begin{align}
C^{-1}_{Ii} = (v_I - \varepsilon_i) \prod_{\substack{K \neq I  \\ k \neq i} } \frac{(\varepsilon_i - v_K)(v_I - \varepsilon_k)}{(v_I - v_K)(\varepsilon_i - \varepsilon_k)}
\end{align}
but this expression is not optimal numerically as floating point precision will be lost rather quickly. As a result, $C$ will be inverted numerically. Upper case letters have been used to label parameters in the ABA basis. Inverting \eqref{eq:trans} gives
\begin{align}
\textbf{a}^{\dagger}_{\sigma} = (C^{\sigma})^{-1} a^{\dagger}(\textbf{v}^{\sigma}),
\end{align}
so that the Coulomb Hamiltonian can then be written in the ABA parameters
\begin{align}
\hat{H}_C = \sum_{IJ} \sum_{\sigma} \tilde{h}^{\sigma}_{IJ} a^{\dagger}(v^{\sigma}_I) a(v^{\sigma}_J) 
+ \frac{1}{2} \sum_{IJKL} \sum_{\sigma\tau} \tilde{V}^{\sigma\tau}_{IJKL} a^{\dagger}(v^{\sigma}_I) a^{\dagger}(v^{\tau}_J) a(v^{\tau}_L) a(v^{\sigma}_K)
\end{align}
where the integrals have been transformed
\begin{align}
\tilde{h}^{\sigma}_{IJ} &= \sum_{ij} h_{ij} (C^{\sigma}_{Ii})^{-1} (C^{\sigma}_{Jj})^{-1} \\
\tilde{V}^{\sigma\tau}_{IJKL} &= \sum_{ijkl} V_{ijkl} (C^{\sigma}_{Ii})^{-1} (C^{\tau}_{Jj})^{-1} (C^{\sigma}_{Kk})^{-1} (C^{\tau}_{Ll})^{-1}.
\end{align}

Excited ABA states are generated by acting on the ground state with the ABA operators. Specifically the single excitation from the $a$th occupied orbital with spin $\mu$ to the $p$th virtual with spin $\nu$ is written
\begin{align}
a^{\dagger}(v^{\nu}_p) a(v^{\mu}_a) \ket{ \{v\}} = \frac{\partial A(v^{\mu}_a)}{\partial v^{\mu}_a} \ket{ \{v\}^{p\nu}_{a\mu}}
\end{align}
and likewise double excitations are generated
\begin{align}
a^{\dagger}(v^{\kappa}_p) a^{\dagger}(v^{\eta}_q) a(v^{\nu}_b)a(v^{\mu}_a) \ket{\{v\}} = 
\frac{\partial A(v^{\mu}_a)}{\partial v^{\mu}_a} \frac{\partial A(v^{\nu}_b)}{\partial v^{\nu}_b} \ket{ \{v\}^{p\kappa q\eta}_{a\mu b\nu } }.
\end{align}

Again, the 2nd order energy correction will have contributions from singles and doubles
\begin{align}
E^{(2)}_{U,0} &= \sum_{\substack{a \in occ \\ p \in virt} } \sum_{\mu\nu} \frac{1}{v^{\mu}_a - v^{\nu}_p} \frac{\vert \braket{ \{v\} | \hat{H}_C | \{v\}^{p\nu}_{a\mu}} \vert^2}{\braket{\{v\}|\{v\}}\braket{\{v\}^{p\nu}_{a\mu}| \{v\}^{p\nu}_{a\mu} }} \nonumber \\
&+ \frac{1}{4} \sum_{\substack{ab \in occ \\ pq \in virt} } \sum_{\mu\nu\kappa\eta} \frac{1}{v^{\mu}_a + v^{\nu}_b - v^{\kappa}_p - v^{\eta}_q}
\frac{\vert \braket{ \{v\} | \hat{H}_C | \{v\}^{p\kappa q\eta}_{a\mu b\nu}} \vert^2}{\braket{\{v\}|\{v\}}\braket{\{v\}^{p\kappa q\eta}_{a\mu b\nu}| \{v\}^{p\kappa q\eta}_{a\mu b\nu} }}
\end{align}
for which we will have to work out TDM elements. This may be done by Wick's theorem as the ABA operators yield orbitals that are orthogonal but not normal. The coupling between the ABA ground state and single excitations has a contribution from 1-electron operators:
\begin{align}
\braket{ \{v\} | a^{\dagger}(v^{\sigma}_I) a(v^{\sigma}_J) | \{v\}^{p\nu}_{a\mu} } = \delta_{aI}\delta_{pJ}\delta_{\nu\sigma}\delta_{\mu\sigma}
\frac{\partial A(v^{\nu}_p)}{\partial v^{\nu}_p} \braket{ \{v\} | \{v\} }
\end{align}
and a more complicated contribution from 2-electron operators
\begin{align}
&\braket{ \{v\} | a^{\dagger}(v^{\sigma}_I) a^{\dagger}(v^{\tau}_J) a(v^{\tau}_L) a(v^{\sigma}_K) | \{v\}^{p\nu}_{a\mu}} \nonumber \\
&= \frac{\partial A(v^{\nu}_p)}{\partial v^{\nu}_p} \frac{\partial A(v^{\tau}_L)}{\partial v^{\tau}_L} \braket{\{v\} | \{v\}} \delta_{pK} \delta_{JL} \delta_{aI}\delta_{\nu\sigma}\delta_{\mu\sigma} \sum_{c\in occ}\delta_{cJ} \nonumber \\
&+ \frac{\partial A(v^{\nu}_p)}{\partial v^{\nu}_p} \frac{\partial A(v^{\sigma}_K )}{\partial v^{\sigma}_K} \braket{ \{v\} | \{v\} } \delta_{pL}\delta_{IK}\delta_{aJ}\delta_{\nu\tau}\delta_{\mu\tau} \sum_{c \in occ} \delta_{cI} \nonumber \\
&- \frac{\partial A(v^{\nu}_p)}{\partial v^{\nu}_p} \frac{\partial A(v^{\sigma}_L)}{\partial v^{\sigma}_L} \braket{ \{v\} | \{v\} } \delta_{pK}\delta_{IL}\delta_{aJ}\delta_{\nu\sigma}\delta_{\mu\tau}\delta_{\sigma\tau} \sum_{c\in occ} \delta_{cI} \nonumber \\
&- \frac{\partial A(v^{\nu}_p)}{\partial v^{\nu}_p} \frac{\partial A(v^{\sigma}_K)}{\partial v^{\sigma}_K} \braket{ \{v\} | \{v\} } \delta_{pL}\delta_{JK}\delta_{aI}\delta_{\nu\tau}\delta_{\mu\sigma}\delta_{\sigma\tau} \sum_{c\in occ} \delta_{cJ}.
\end{align}

The only non-vanishing coupling between the ABA ground state and double excitations is
\begin{align}
\braket{ \{v\} | a^{\dagger}(v^{\sigma}_I) a^{\dagger}(v^{\tau}_J) a(v^{\tau}_L) a(v^{\sigma}_K)| \{v\}^{p\kappa q\eta}_{a\mu b\nu} } = 
\frac{\partial A(v^{\kappa}_p)}{\partial v^{\kappa}_p} \frac{\partial A(v^{\eta}_q)}{\partial v^{\eta}_q} \braket{ \{v\} | \{v\}} \times \nonumber \\
 \times ( \delta_{aI}\delta_{bJ}\delta_{\mu\sigma}\delta_{\nu\tau}  - \delta_{aJ}\delta_{bI}\delta_{\mu\tau}\delta_{\nu\sigma}  )
 ( \delta_{pK} \delta_{qL} \delta_{\kappa\sigma}\delta_{\eta\tau} - \delta_{pL}\delta_{qK}\delta_{\kappa\tau}\delta_{\eta\sigma} ) .
\end{align}
The final expression for the unrestricted 2nd order correction is thus
\begin{align}
E^{(2)}_{U,0} &= \sum_{\substack{a \in occ \\ p \in virt} } \sum_{\mu} \frac{1}{v^{\mu}_a - v^{\mu}_p} \frac{\partial A(v^{\mu}_p)}{\partial v^{\mu}_p} \frac{\partial A(v^{\mu}_a)}{\partial v^{\mu}_a} ( \tilde{D}^{p\mu}_{a\mu})^2 \nonumber \\
&+ \frac{1}{2} \sum_{\substack{ab \in occ \\ pq \in virt} } \sum_{\mu\nu} 
\frac{\tilde{V}^{\mu\nu}_{pqab} \tilde{V}^{\mu\nu}_{pqab}}{v^{\mu}_a + v^{\nu}_b - v^{\mu}_p - v^{\nu}_q} 
\frac{\partial A(v^{\mu}_p)}{\partial v^{\mu}_p} \frac{\partial A(v^{\nu}_q)}{\partial v^{\nu}_q} \frac{\partial A(v^{\mu}_a)}{\partial v^{\mu}_a} \frac{\partial A(v^{\nu}_b)}{\partial v^{\nu}_b} \nonumber \\
&- \frac{1}{2} \sum_{\substack{ab \in occ \\ pq \in virt} } \sum_{\mu}
\frac{\tilde{V}^{\mu\mu}_{pqab} \tilde{V}^{\mu\mu}_{pqba}}{v^{\mu}_a + v^{\mu}_b - v^{\mu}_p - v^{\mu}_q}
\frac{\partial A(v^{\mu}_p)}{\partial v^{\mu}_p} \frac{\partial A(v^{\mu}_q)}{\partial v^{\mu}_q} \frac{\partial A(v^{\mu}_a)}{\partial v^{\mu}_a} \frac{\partial A(v^{\mu}_b)}{\partial v^{\mu}_b}
\end{align}
where the singles contribution is
\begin{align}
\tilde{D}^{p\mu}_{a\mu} = \tilde{h}^{\mu}_{pa} + \sum_{c\in occ} \left(
\sum_{\sigma} \tilde{V}^{\sigma\mu}_{cpca}\frac{\partial A(v^{\sigma}_c)}{\partial v^{\sigma}_c} - \tilde{V}^{\mu\mu}_{cpac} \frac{\partial A(v^{\mu}_c)}{\partial v^{\mu}_c}
\right).
\end{align}

Taking the restricted limit, we arrive at
\begin{align} \label{eq:resPT2}
E^{(2)}_{R,0} &= 2 \sum_{\substack{a \in occ \\ p \in virt} } \frac{1}{v_a - v_p} \frac{\partial A(v_p)}{\partial v_p} \frac{\partial A(v_a)}{\partial v_a}
\left(
\tilde{h}_{pa} + \sum_{c\in occ} (2\tilde{V}_{cpca} - \tilde{V}_{cpac}) \frac{\partial A(v_c)}{\partial v_c}
\right)^2 \nonumber \\
&+ \sum_{\substack{ab \in occ \\ pq \in virt} } \frac{(2\tilde{V}_{pqab} - \tilde{V}_{pqba}) \tilde{V}_{pqab}}{v_a + v_b - v_p - v_q}
\frac{\partial A(v_p)}{\partial v_p} \frac{\partial A(v_q)}{\partial v_q} \frac{\partial A(v_a)}{\partial v_a} \frac{\partial A(v_b)}{\partial v_b}.
\end{align}
Equation \eqref{eq:resPT2} is quite similar to \eqref{eq:resPT1}, except that the integrals have been transformed in the opposite manner.

\section{Discussion}
We derived two expressions, \eqref{eq:resPT1} and \eqref{eq:resPT2}, for the restricted 2nd order RSPT energy correction. Both were implemented, and found to give numerically the same results. Unfortunately these results were ambiguous.  The ABA parameters $\{\varepsilon\},g$ may be changed without affecting the mean-field energy \eqref{eq:ABA_R}, though they have a large effect on the 2nd-order energy. In addition, unlike the case for MP2, the singles couple to the ground state through $\hat{H}_C$. To study these effects we tried minimizing the energy functional
\begin{align}
F = E^{ABA}_R +  \sum_{\substack{a \in occ \\ p \in virt} } \vert \braket{\{v\} | \hat{H}_C | \{v\}^p_a} \vert^2
\end{align}
which leads to numerically the same mean-field energy, and the singles do vanish. \emph{But so do the doubles} and thus the 2nd order RSPT correction is numerically zero. The $N+1$ variables $\{\varepsilon\},g$ are enough to reproduce the ground state RHF energy as all that is required is an idempotent 1-RDM in the correct basis. However, they are not enough to reproduce the entire RHF spectrum and hence do not reproduce the MP2 perturbative correction. To fix this, one could employ a Hylleraas functional\cite{hylleraas} to solve for the perturbative correction variationally. Solving the equations explicitly  on paper would of course lead to the same expressions \eqref{eq:resPT1} and \eqref{eq:resPT2}. Instead one should consider variationally solving for the coefficients $\{c_{ki}\}$
\begin{align}
\ket{\Psi^{(1)}_k} = \sum_i c_{ki} \ket{\Psi^{(0)}_i}
\end{align}
but this amounts to the same number of parameters as an orbital optimization. Including an orbital optimization would necessarily reproduce MP2 as the RHF solution is defined by the optimal orbitals. Similarly, we could solve the CI singles problem for ABA states as a reference wavefunction, but again this amounts to employing $N^2$ parameters. Therefore we do not consider it productive to demonstrate numerically.

\section{Conclusion}
Perturbative corrections for the ABA for individual electrons were calculated. The first expression \eqref{eq:resPT1} was computed in the primitive basis using TDM elements obtained from the ABA. The second expression \eqref{eq:resPT2} was computed in the ABA basis using particle-hole excitations. Both expressions are in close analogy with MP2 corrections to RHF. Numerically computing these corrections lead to ambiguous results that could, in principle, be corrected with an orbital optimization or a CI singles solution. Going forward to pairs of electrons, we understand that an orbital optimization is required along with the ABA wavefunction. Care will be required when computing perturbative corrections.

\begin{acknowledgments}
We are grateful for support form NSERC. This research was enabled by support from CalculQu\'{e}bec and Compute Canada. 
\end{acknowledgments}

\bibliography{BAHF_PT_arXiv}

\bibliographystyle{unsrt}

\end{document}